\global\def\draftcontrol{0}

   \def\versionno{NP GV}

\catcode`\@=11

\expandafter\ifx\csname draftcontrol\endcsname\relax\global\def\draftcontrol{0} 
\fi 

{\count255=\time\divide\count255 by 60 
\xdef\hourmin{\number\count255} 
\multiply\count255 by-60\advance\count255 by\time 
\xdef\hourmin{\hourmin:\ifnum\count255<10 0\fi\the\count255}} 
\def\draftdate{\number\month/\number\day/\number\year\ \ \ \hourmin } 

\newcommand\makepapertitle{\par

  \begingroup 
    \renewcommand\thefootnote{\@fnsymbol\c@footnote}%
    \def\@makefnmark{\rlap{\@textsuperscript{\normalfont\@thefnmark}}}%
    \long\def\@makefntext##1{\parindent 1em\noindent 
            \hb@xt@1.8em{%
                \hss\@textsuperscript{\normalfont\@thefnmark}}##1}%
     \newpage 
     \global\@topnum\z@   
     \@makepapertitle 
     \thispagestyle{empty}\@thanks 
  \endgroup 
  \setcounter{footnote}{0}%
  \global\let\thanks\relax 
  \global\let\makepapertitle\relax 
  \global\let\@makepapertitle\relax 
  \global\let\@thanks\@empty 
  \global\let\@author\@empty 
  \global\let\@date\@empty 
  \global\let\@title\@empty 
  \global\let\title\relax 
  \global\let\author\relax 
  \global\let\date\relax 
  \global\let\and\relax 
  \def\version{\let\version\@version\@gobble} 
} 
\def\@makepapertitle{%
  \newpage 
   \ifnum\draftcontrol=1 {} 
   \version\versionno 
   \vskip 5.5em%
   \else 
   \hfill\hbox to 3cm {\parbox{4.5cm}{\@pubnum}\hss}%
   \vskip 6.5em%
   \fi 
   \begin{center}%
   \let \footnote \thanks 
      {\hskip -0\textwidth \hbox to 1\textwidth%
        {\centerline{\Large\bf{\noindent\@title}}}}%
     \vskip 2em%
     {\normalsize
       \lineskip .5em%
       \begin{tabular}[t]{c}%
         \@author 
       \end{tabular}\par}%
     \vskip 1.5em%
     {\@bstract}%
     \end{center}%
     \vfill
     \@date%
     \vskip 1.5em%
   \par 
} 

\gdef\@pubnum{} 
\def\pubnum#1{%
  \gdef\@pubnum{#1}} 

\gdef\@bstract{} 
\def\Abstract#1{%
  \gdef\@bstract{%
   \parbox{\textwidth-0pc}{%
   \centerline{\bf Abstract}\penalty1000 
   \noindent
   \renewcommand\baselinestretch{1.0} 
   {#1}}} 
} 

\gdef\@email{}
\def\email#1{%
   \gdef\@email{%
   Email: {\tt #1}}
}

\def\ps@paper{\let\@mkboth\@gobbletwo%
     \ifnum\draftcontrol=1 
        \def\@oddfoot{\hbox to \textwidth{\tiny \versionno \hfil\tiny\draftdate}%
        \hskip -\textwidth \hbox to \textwidth{\hfil\rm\thepage\hfil}}%
     \else\def\@oddfoot{\hbox to \textwidth{\hfil\rm\thepage\hfil}} 
     \fi 
     \let\@evenfoot\@oddfoot 
} 

\def\body{\clearpage 
          \pagestyle{paper} 
        } 
\newenvironment{acknowledgments}{%
\vskip 3.25ex 
\addcontentsline{toc}{section}{Acknowledgments}
\noindent {\bf Acknowledgments} 
} 

\def\@version#1{\ifnum\draftcontrol=1 
\typeout{}\typeout{#1}\typeout{} 
\vskip3mm\centerline{\hbox{\fbox{\normalsize{\tt DRAFT -- #1 -- } 
                   {\draftdate}}}}\vskip3mm 
\fi} 
\let\version\@version 
\long\def\eqlabel#1{\ifnum\draftcontrol=1 
                    \tag@false  
                    \tag*{(\theequation) \hbox to -0.2cm{\hspace{0cm}\small{#1}\hss}} 
                    \refstepcounter{equation}  
                    \edef\@currentlabel{\theequation} 
                    \ltx@label{#1}          
                    \else 
                    \label{#1} 
                    \fi 
                    } 
\let\st@bibitem\@bibitem 
\let\st@lbibitem\@lbibitem 
\ifnum\draftcontrol=1 
  \def\@bibitem#1{%
    \st@bibitem{#1}\a@@label{#1}\ignorespaces} 
  \def\@lbibitem[#1]#2{%
    \st@lbibitem[#1]{#2}\a@@label{#2}\ignorespaces} 
  \def\a@@label#1{%
    \gdef\a@lab{\smash{\normalfont\small#1}} 
    \ifvmode 
      \if@inlabel 
        \global\setbox\@labels\hbox{%
          \llap{\a@lab\let\a@lab\relax 
                \kern\@totalleftmargin\kern\marginparsep}%
          \box\@labels}%
      \fi 
    \fi} 
\fi 

\documentclass[12pt,letterpaper]{article} 

\usepackage{amsmath,bm,amsfonts,amssymb,array,calc,amsthm,rotating,amscd}
\usepackage{epsfig,psfrag} 
\usepackage{rotating}
\usepackage{amscd}
\usepackage{graphicx}
\usepackage{color}
\usepackage[colorlinks=false]{hyperref}

\renewcommand\baselinestretch{1.25} 
\renewcommand\section{\@startsection {section}{1}{\z@}%
                                   {-3.5ex \@plus -1ex \@minus -.2ex}%
                                   {2.3ex \@plus.2ex}%
                                   {\normalfont\large\bfseries}} 
\renewcommand\subsection{\@startsection{subsection}{2}{\z@}%
                                   {-3.25ex\@plus -1ex \@minus -.2ex}%
                                   {1.5ex \@plus .2ex}%
                                   {\normalfont\normalsize\bfseries}} 
\renewcommand\subsubsection{\@startsection{subsubsection}{3}{\z@}%
                                   {-3.25ex\@plus -1ex \@minus -.2ex}%
                                   {1.5ex \@plus .2ex}%
                                   {\normalfont\normalsize\it}} 
\renewcommand\paragraph{\@startsection{paragraph}{4}{\z@}%
                                   {-1.75ex\@plus -1ex \@minus -.2ex}%
                                   {1ex \@plus .2ex}%
                                   {\normalfont\normalsize\bf}} 
\renewcommand\subparagraph{\@startsection{subparagraph}{5}{\z@}%
                                   {-1.25ex\@plus -0ex \@minus -.2ex}%
                                   {-2ex \@plus .2ex}%
                                   {\normalfont\normalsize\it}}

\numberwithin{equation}{section}

\long\def\@makecaption#1#2{%
  \vskip\abovecaptionskip
  \sbox\@tempboxa{{\bf #1:} #2}%
  \ifdim \wd\@tempboxa >\hsize
    {\small\bf #1:} {\small #2}\par
  \else
    \global \@minipagefalse
    \hb@xt@\hsize{\hfil\box\@tempboxa\hfil}%
  \fi
  \vskip\belowcaptionskip}

\renewcommand*\l@section[2]{%
  \ifnum \c@tocdepth >\z@
    \addpenalty\@secpenalty
    \addvspace{.5em \@plus\p@}%
    \setlength\@tempdima{1.5em}%
    \begingroup
      \parindent \z@ \rightskip \@pnumwidth
      \parfillskip -\@pnumwidth
      \leavevmode \bfseries
      \advance\leftskip\@tempdima
      \hskip -\leftskip
      #1\nobreak\hfil \nobreak\hb@xt@\@pnumwidth{\hss #2}\par
    \endgroup
  \fi}
\renewcommand*\l@subsection{\addvspace{.0em \@plus\p@}\@dottedtocline{2}{1.5em}{2.3em}}
\renewcommand*\l@subsubsection{\addvspace{-.2em \@plus\p@}\@dottedtocline{3}{3.8em}{3.2em}}

\def\hepth#1{\href{http://xxx.arxiv.org/abs/hep-th/#1}{{arXiv:hep-th/#1}}}

\def\arxiv#1#2{\href{http://xxx.arxiv.org/abs/#1}{{arXiv:#1 [#2]}}}

\definecolor{refcol}{rgb}{0.2,0.2,0.8}
\definecolor{eqcol}{rgb}{.6,0,0}
\definecolor{purple}{cmyk}{0,1,0,0}

\gdef\@citecolor{refcol}
\gdef\@linkcolor{eqcol}
\def\colorlinkspurple{\gdef\@urlcolor{purple}}
\def\colorlinksblue{\gdef\@urlcolor{blue}}
\def\colorlinksred{\gdef\@urlcolor{red}}

\def\ie{{\it i.e.}}

\def\cf{{\it cf.}}

\def\revise#1       {\raisebox{-0em}{\rule{3pt}{1em}}%
                     \marginpar{\raisebox{.5em}{\vrule width3pt\ 
                     \vrule width0pt height 0pt depth0.5em 
                     \hbox to 0cm{\hspace{0cm}{%
                     \parbox[t]{4em}{\raggedright\footnotesize{#1}}}\hss}}}}

\def\ii           {{\it i}}

\def\Re           {{\rm Re\hskip0.1em}}

\def\sqr#1#2{{\vcenter{\vbox{\hrule height.#2pt   
 \hbox{\vrule width.#2pt height#1pt \kern#1pt 
 \vrule width.#2pt}\hrule height.#2pt}}}}

\newcommand{\N}{\mathbb N}
\renewcommand{\S}{\mathbb S}

\newcommand{\Fcal}{\mathcal F}

\newcommand{\Lcal}{\mathcal L}
\newcommand{\Ccal}{\mathcal C}

\newcommand{\Ncal}{\mathcal N}

\newcommand{\Mcal}{\mathcal M}

\newcommand{\ep}{\epsilon}

\newcommand{\beq}{\begin{equation}}
\newcommand{\eq}{\end{equation}}
\newcommand{\req}[1]{(\ref{#1})}

\catcode`\@=12 

\begin{document} 


\title{Mellin-Barnes Representation of the Topological String}

\pubnum{
SNUTP15-006
}
\date{August 2015}

\author{
Daniel Krefl$^a$ \\[0.2cm]
\it  $^a$ Center for Theoretical Physics, SNU, Seoul, South Korea\\
}

\Abstract{
We invoke integrals of Mellin-Barnes type to analytically continue the Gopakumar-Vafa resummation of the topological string free energy in the string coupling constant, leading to additional non-perturbative terms. We also discuss in a similar manner the refined and Nekrasov-Shatashvili limit version thereof. The derivation is straight-forward and essentially boils down to taking residue. This allows us to confirm some related conjectures in the literature at tree-level.
}

\makepapertitle

\body

\version\versionno

\vskip 1em



\section{Introduction}
Recently, the question of a possible non-perturbative completion of the topological string (and related theories) received renewed interest. This has been triggered by three a priori independent developments. Firstly, via progress in translating and applying well-established techniques like resurgence and exact quantization to the topological string context, see in particular \cite{ASV11,SESV13, K13, K14, BD15}. Secondly, via insight gained from the study of ABJM theories in the Fermi gas formalism, quantization thereof, and the resulting spectral determinants, particularly \cite{HMO12,HMMO13}. Thirdly, via the connection between $\Ncal=1$ superconformal theories on $\S^5$ and the refined topological string \cite{LV12}. 

The latter two approaches led to a proposal for a non-perturbative completion of topological string free energies, at least in a particular regime in K\"ahler moduli space. More precisely, the authors of \cite{HMO12} conjectured the general non-perturbative completion of the topological string on a toric Calabi-Yau \cite{HMMO13} to be essentially encoded in the corresponding Nekrasov-Shatashvili free energies. The support for the validity of their conjecture mainly comes so far from numerics (but has been analytically confirmed for the resolved conifold in \cite{KM15}). The proposal of \cite{LV12} is more general, as they propose the non-perturbative completion of the refined topological string (in Gopakumar-Vafa form), which includes the ordinary topological string under a suitable specialization of equivariant parameters. The approach of \cite{LV12} is based on making use of analytic properties of multiple-sine functions and the connection to superconformal theories, implying a triple factor structure for the refined partition functions. Under specialization of equivariant parameters their proposal agrees with \cite{HMMO13} (modulo some differences in choice of K\"ahler class and flat coordinates). The precise relation of the latter two approaches and the implications for the resurgence based ones mentioned above first is at the time being not clear and remains an avenue for future research. However, consistency requires that in overlapping regimes of validity they should yield the same result.

The purpose of this note is to analytically confirm the conjectured non-perturbative terms of \cite{HMMO13} and \cite{LV12} for the tree-level part of the partition functions in a very straight-forward and simple way. In essence, the analytic structure of the Gopakumar-Vafa free energy \cite{GV98a,GV98b} under analytic continuation in the string coupling constant, \ie, the pole structure in the complex plane, \cf, \cite{HMO12}, determines a minimal necessary non-perturbative sector. In spirit closest to our approach below comes \cite{H15}, though there are some essential differences. 

The remaining non-trivial claim of the conjectures is therefore that the particular non-perturbative structure of the partition function present at tree-level generally persists to higher genus, on which, unfortunately, we can not give a conclusive answer with the formalism to be introduced in the remaining part of this introduction. 

The main idea we present in this note is readily explained. Recall the Mellin-Barnes integral representation of the hypergeometric function
\beq
_2 F_1(a,b,c;z) = \frac{\Gamma(c)}{2\pi\ii \Gamma(a)\Gamma(b)}\int_\Ccal dx\, \frac{\Gamma(a+x)\Gamma(b+x)\Gamma(-x)}{\Gamma(c+x)} (-z)^x\,,
\eq
convergent for $|\arg(-z)|<\pi$ (see for instance \cite{PK01}). The integration contour $\Ccal$ runs (upward) along the imaginary axis, suitably indented to separate the poles of $\Gamma(-x)$ from the poles of $\Gamma(a+x)\Gamma(b+x)$. It can be shown that the contour can be closed, picking up either the poles of $\Gamma(-x)$ along the positive real axis or the poles of $\Gamma(a+x)\Gamma(b+x)$. We denote the former deformed contour as $\Ccal_+$ (running clockwise, requiring that $|z|<1$), while the latter as $\Ccal_-$ (running anti-clockwise, leading to the asymptotic expansion for large $|z|$).

In particular, from the above integral representation it follows that 
$$
\log(1+z) = z\,\,  _2 F_1(1,1,2;-z)\,,
$$
and so 
\beq\eqlabel{LogIntRep}
\log(1+z)= \frac{1}{2\pi\ii }\int_{\Ccal_+} dx\, \frac{\Gamma^2(1+x)\Gamma(-x)}{\Gamma(2+x)}\, z^{1+x}\,.
\eq
Indeed, deforming to $\Ccal_+$ in order to pick up the simple poles of $\Gamma(-x)$, we recover via residue taking the series expansion of $\log(1+z)$ from the integral representation \req{LogIntRep}. We can also close the contour on the other side, \ie, $\Ccal_-$, and recover again
$\log(1+z)$. However, we will not make any further use of $\Ccal_-$ in this note.

The fact we will make use of is that a significant class of topological string free energies $\Fcal$ at the large volume point in K\"ahler moduli space can be expressed at least in part qualitatively as summation
\beq\eqlabel{Fqualitative}
\Fcal \sim \sum_{{\bf n}>0} \log(1+{\bf z^n}\, Q)\,,
\eq
where ${z_i}$ parameterize the coupling constant(s) and $Q$ the K\"ahler parameters. (Bold letters are understood as ${\bf x}^{\bf n}=x_1^{n_1} x_2^{n_2}\dots$ .) Essentially, this corresponds to the Donaldson-Thomas form of the free energy. 

We can sum over the integral representation \req{LogIntRep} making use of the geometric series (for $\Re z_i < 1$ understood as analytic continuation via Borel resummation) 
$$
\sum_{{\bf n} >0} ({\bf z}^{1+x})^{{\bf n}}=\prod_{i} \frac{1}{1-z_i^{1+x}}\,,
$$
effectively yielding a resummation and analytic continuation of $\Fcal$. Note that in the process of resummation we may pick up ``new'' poles at points $x_i^*$ for which
$$
z_i^{1+x_i^*} = 1\,,
$$
holds. As we will see below, these poles encode non-perturbative information and are therefore also referred to as ``non-perturbative'' poles. Hence, for free energies taking a form similar to \req{Fqualitative}, we are able to easily analytically continue in the parameters ${\bf z}$, thereby arriving at a trans-series like expansion, which goes beyond the usual weak coupling analysis.

We believe that the Mellin-Barnes method sketched above will be of use more generally in the study of partition functions occuring in the context of supersymmetric theories in various dimensions. For instance, the logarithm of superconformal indices often takes a structure similar to \req{Fqualitative}, \cf, \cite{BBMR08}.

The outline is as follows. In the next section we will discuss the usual Gopakumar-Vafa expansion of the topological string free energy, thereby confirming the conjecture of \cite{HMMO13} at tree-level. However, due to absence of non-perturbative poles at higher genus, we can not confirm via our approach beyond tree-level. In section \ref{RefGVsec} the refined case will be considered. Here, things will be slightly different and we have a surviving non-perturbative sector beyond tree-level, under certain conditions on the equivariant parameters. We will also discuss the Nekrasov-Shatashvili limit in this section. We conclude in section \ref{ExampleSec} with an explicit and non-trivial example, illustrating that the Mellin-Barnes method introduced above is wider applicable. In detail, the method can be easily applied to Chern-Simons theory on $\S^3$, thereby rederiving some of the results of \cite{KM15} in a far simpler way (the underlying reason being uniqueness of analytic continuation).

\section{Gopakumar-Vafa expansion}

\paragraph{$g=0$}
\label{GVsecg0}
Recall that we know since Gopakumar-Vafa \cite{GV98a,GV98b} that the genus zero part of the topological string partition function takes the form 
\beq\eqlabel{Z0def}
Z^{(0)}=M_{g_s}(1)^{-\frac{\chi}{2}}\prod_{\mathbf d} M_{g_s}(Q^{\mathbf d})^{n_{\mathbf d}^{(0)}}\,,
\eq
where $\chi$ is the Euler characteristic of the Calabi-Yau, $\mathbf d$ denotes the K\"ahler class, $n^{(0)}_{\mathbf d}$ the tree-level Gopakumar-Vafa invariants, and the function $M_{g_s}(Q)$ is defined as
$$
M_{g_s}(Q):=\prod_{n=1}^\infty \left(1- q^n Q\right)^n\,.
$$
We further defined $q:=e^{-2\pi\ii\, g_s}$ and $Q:=e^{-2\pi\ii \,\mathbf t}$ with $\mathbf t$ the K\"ahler parameters and $g_s$ the string coupling constant.

Note that $M_{g_s}(1)^{-1}$ is the MacMohan function, the generating function for 3d partitions. Here it encodes the constant map contribution. The fact that one can express the topological string partition function at genus zero via the function $M_{g_s}$, as in \req{Z0def}, is central for the duality between topological strings and classical crystals at large $g_s$ \cite{ORV03,INOV03}, which led to the celebrated Donaldson-Thomas / Gromov-Witten correspondence. In fact, expanding the Donaldson-Thomas like partition function \req{Z0def} for small $g_s$, the usual Gopakumar-Vafa resummation of the topological string free energy is easily recovered. 

However, what we will show in detail below is that even if one does not expand $Z^{(0)}$ for small $g_s$, one can still rewrite the corresponding free energy $\Fcal:=\log Z$ as a Gopakumar-Vafa like summation, but with an additional non-perturbative sector. The resulting trans-series like expansion matches the conjectured form of \cite{HMMO13,LV12}, and in the case of the conifold the independent calculation of \cite{KM15} via universal Chern-Simons theory.

From the integral representation \req{LogIntRep} it follows via a simple resummation that
\beq\eqlabel{MintRep}
\Mcal_{g_s}(Q) = \frac{1}{8\pi\ii} \int_{\Ccal_+} dx \, \frac{\Gamma^2(1+x)\Gamma(-x)}{\sin^2\left(\pi(x+1)g_s\right)\Gamma(2+x)} \, (-Q)^{1+x} \,,
\eq
where we defined $\Mcal:=\log M$. As we can transform the $\sin^2$ via Euler's reflection formula $\Gamma(1-z)\Gamma(z)=\frac{\pi}{\sin\pi z}$ to Gamma functions, the above integral representation for $\Mcal_q(Q)$ is of Mellin-Barnes type. The integrand picks up besides the ray of simple poles at
$$
x_p^* = n\,,
$$
a ray of order 2 poles at
$$
x_{np}^*= -1+\,\frac{n+1}{g_s}\,,
$$
with $n\in \mathbb N$. For reasons becoming more clear below we will refer to the former poles as {\it perturbative} and the latter as {\it non-perturbative}. Therefore, we split $\Mcal$ into two parts, \ie,
\beq\label{Msplit}
\Mcal = \Mcal^p+\Mcal^{np}\,.
\eq
The contribution of the perturbative poles can be readily inferred, yielding
$$
\Mcal^p_{g_s}(Q) = \frac{1}{4}\sum_{n=1}^\infty \frac{Q^{n}}{n \sin^2\left(\pi n\, g_s\right)}\,.
$$

It remains to discuss the non-perturbative poles. Note  that for functions $f(z)$ regular and $g(z)$ having an order 2 pole at $z=a$, we have
\beq\eqlabel{order2residue}
{\rm Res}_{z=a} \frac{f(z)}{g(z)} = \frac{6f'(a) g''(a)-2f(a) g'''(a)}{3(g''(a))^2}\,. 
\eq
Straight-forward calculation then yields,
\beq\eqlabel{MnpgsExpansion}
\Mcal^{np}_{g_s}(Q)=-\frac{1}{4 \pi}\sum_{n=1}^\infty \frac{(-Q)^{\frac{n}{g_s}}}{n^2 \sin\left(\frac{\pi n}{g_s}\right)} \Big(1-\frac{n}{g_s}\log(-Q) +\frac{n\pi}{g_s} \cot\left(\frac{n \pi }{g_s}\right) \Big)  \,.
\eq

Let us consider the general genus zero free energy given by \req{Z0def}, \ie,
$$
\Fcal^{(0)} =-\frac{\chi}{2} \Mcal_{g_s}(1)+ \sum_{\mathbf d} n_{\mathbf d}^{(0)} \Mcal_{g_s}(Q^{\mathbf d})\,.
$$
We immediately deduce that we have in general
\beq
\boxed{
\Fcal^{(0)}=-\frac{\chi}{2}\left(\Mcal^p_{g_s}(1)+\Mcal^{np}_{g_s}(1)\right)+\sum_{\mathbf d} n_{\mathbf d}^{(0)}\left(\Mcal^p_{g_s}(Q^{\mathbf d})+\Mcal^{np}_{g_s}(Q^{\mathbf d})\right)
}\,.
\eq
We will come back to this expression in section \ref{RefGVsec}.

\paragraph{$g>0$}
One can as well write the higher genus part of the topological string partition function (at large volume) in a similar product form as the genus zero part given in \req{Z0def}, see for instance \cite{KKRS04,DM07}. Namely, defining
$$
L^{(g>0)}_{g_s}(Q):=\prod_{l=0}^{2g-2} (1-q^{g-l-1} Q)^{\frac{(-1)^{g+l}}{l!(2g-2-l)!}}\,,
$$
we can write
\beq\label{DefZgg0}
Z^{(g>0)}=\prod_{\mathbf d}\prod_{g=1}^{\infty} L^{(g)}_{g_s}(Q^{\bf d})^{(2g-2)!\, n_{\mathbf d}^{(g)}}\,,
\eq
where $n_{\mathbf d}^{(g)}$ are the Gopakumar-Vafa invariants of higher genus. Defining $\Lcal^{(g)}:=\log L^{(g)}$, using again the integral representation \req{LogIntRep} and summing over $l$, we infer
\beq\label{LgintRep}
\Lcal^{(g>0)}_{g_s}(Q) =\,  \frac{ 2^{2g-2}(-1)^{g-1}}{2\pi\ii\,\Gamma(2g-1)} \int_{\Ccal_{+}} dx \,\frac{\sin^{2g-2}\left(\pi(x+1)g_s\right)\Gamma^2(1+x)\Gamma(-x)}{\Gamma(2+x)}\, (-Q)^{1+x}\,.
\eq
Taking residue yields
$$
\Lcal^{(g>0)}_{g_s}(Q) = \frac{2^{2g-2}(-1)^{g-1}}{(2g-2)!}\sum_{n=1}^\infty \frac{\sin^{2g-2}\left(\pi n\,g_s\right)}{n} \, Q^{n} \,.
$$
We infer via \req{DefZgg0} the well-known (perturbative) Gopakumar-Vafa expansion
$$
\Fcal^{(g>0)}=\sum_{\mathbf d}\sum_{n=1}^\infty\sum_{g=1}^\infty n^{(g)}_{\mathbf d} \frac{\left(2\sin\left(\pi n\, g_s\right)\right)^{2g-2}}{n}\, Q^{n \mathbf d}\,.
$$
However, in contrast to $g=0$, we do not have non-perturbative poles in the integral representation \req{LgintRep}. Therefore, the analytic continuation of the partition function \req{DefZgg0} does not automatically imply a non-perturbative sector at $g>0$, and therefore the general validity of the conjectured non-perturbative structure of \cite{HMMO13} stays elusive in our approach.

One should compare to the related Schwinger-integral based discussion of \cite{PS09}, where the failure to obtain results for $g>0$ has been explained to be due to the ``pole at infinity'' being an essential singularity. Here, we do not see an easy escape route, as \req{LgintRep} is clearly of Mellin-Barnes type (after invoking Euler's reflection formula) and convergent under suitable restrictions on $Q$ (\cf, \cite{PK01}, Lemma 2.5). However, viewing $Q$ as non-perturbative flat coordinate, there is always room for a non-perturbative sector under expansion of $Q$ in terms of the perturbative coordinates.

\section{Refined Gopakumar-Vafa expansion}
\label{RefGVsec}
The refined Gopakumar-Vafa expansion reads 
\beq\eqlabel{RefGV}
\Fcal_{ref}=\sum_{{\bf d}} \sum_{g_L,g_R=0}^\infty \sum_{m=1}^\infty n^{(g_L,g_R)}_{\bf d}\frac{\sin\left(\frac{\pi m\ep_-}{2}\right)^{2g_L}\sin\left(\frac{\pi m\ep_+}{2}\right)^{2g_R}}{4m \sin\left(\pi m\ep_1\right)\sin\left(\pi m\ep_2\right)}\, Q^{m{\bf d}}\,,
\eq
with $\ep_+:=\ep_1+\ep_2$ and $\ep_-:=\ep_1-\ep_2$. (Note that we use the representation basis of \cite{HK10,CKK12} rather than the original one of \cite{HIV03}. Further, we rescaled $\ep_i$ by a factor of $2\pi$.) The infinite product representation of the corresponding partition function can be readily inferred to be given by
$$
Z_{ref} = \prod_{\bf d}\prod_{g_L,g_R=0}^\infty \exp\left(\sum_{m=1}^\infty \frac{\sin\left(\frac{\pi m\ep_-}{2}\right)^{2g_L}\sin\left(\frac{\pi m\ep_+}{2}\right)^{2g_R}}{4m \sin\left(\pi m\ep_1\right)\sin\left(\pi m\ep_2\right)}\, Q^{m{\bf d}} \right)^{n^{(g_L,g_R)}_{\bf d}}\,. 
$$
It remains to bring $\log Z_{ref}$ close to the canonical form \req{Fqualitative}. Using the binomial theorem, we can write
$$
\sin\left(\frac{\pi m \ep}{2}\right)^{2g} = (-4)^{-g}e^{\ii \pi m g\ep}\left(1-e^{-\ii \pi  m\ep}\right)^{2g}=(-4)^{-g}e^{\ii\pi  m g\ep} \sum_{k=0}^{2g} \binom{2g}{k}(-1)^{k} e^{-\ii \pi m k \ep}\,.
$$
Further 
$$
\frac{1}{\sin\left(\pi m\ep \right)}=2\ii \frac{e^{-\ii \pi m\ep}}{1-e^{-2\pi\ii m\ep}}=2\ii e^{-\ii\pi  m\ep}\sum_{k=0}^\infty e^{-2\pi\ii m k \ep}\,,
$$
viewed as analytic continuation as long as $\Re e^{-2\pi\ii m\ep}<1$. Hence,
\beq
\begin{split}
Z_{ref} =& \prod_{\bf d}\prod_{g_L,g_R=0}^\infty\prod_{k_L=0}^{2g_L}\prod_{k_R=0}^{2g_R}\prod_{k_1,k_2=0}^\infty \left(1- (q t)^{\frac{g_L-k_L}{2}}(q/t)^{\frac{g_R-k_R}{2}} q^{-k_1-1/2} t^{k_2+1/2} \,Q^{\bf d}\right)^{N_{\bf d}^{(g_L,g_R)}}\,,
\end{split}
\eq
where we defined $q:=e^{2\pi\ii\ep_1}$, $t:=e^{-2\pi\ii\ep_2}$ and so $e^{\ii\ep_+}=q/t, e^{\ii\ep_-}= q t$, and
$$
N_{\bf d}^{(g_L,g_R)}:=(-1)^{k_L+k_R}(-4)^{1-g_L-g_R}\binom{2g_L}{k_L}\binom{2g_R}{k_R}\, n^{(g_L,g_R)}_{\bf d}\,.
$$

\paragraph{$g_L+g_R=0$}
Let us first consider the case with $g_L=g_R=0$. We define
$$
M_{\ep_1,\ep_2}(Q):= \prod_{k_1,k_2=0}^\infty \left(1-q^{-k_1-1/2} t^{k_2+1/2} \,Q\right)\,,
$$
Taking the logarithm and summing we have
$$
\Mcal_{\ep_1,\ep_2}(Q)=-\frac{1}{8\pi\ii}\int_{\Ccal_+} \frac{\Gamma^2(1+x)\Gamma(-x)}{\sin\left(\pi\ep_1(1+x)\right)\sin\left(\pi\ep_2(1+x)\right)\Gamma(2+x)}(-Q)^{1+x}\,dx\,.
$$
The perturbative poles lead to
$$
\Mcal^p_{\ep_1,\ep_2}(Q) = -\frac{1}{4}\sum_{n=1}^\infty \frac{Q^n}{n \sin\left(\pi n \ep_1\right)\sin\left(\pi n \ep_2\right)}\,,
$$
reproducing the main building block of the refined Gopakumar-Vafa expansion \req{RefGV} at $g_L=g_R=0$, as it should be. In fact, $\Mcal^p_{\ep_1,\ep_2}$ corresponds simply to the perturbative free energy of the refined conifold (without constant map contribution).

For the non-perturbative poles, some more care has to be taken as enhancement to order 2 poles may occur for some of the poles, if $\beta:=-\ep_1/\ep_2 \in \mathbb Q$. Here, we take for simplicity $\beta \not\in \mathbb Q$ such that we have two independent rays of non-perturbative simple poles, \ie,
\beq\eqlabel{refNPpoles}
x^*_{np,1} = -1+\frac{n+1}{\ep_1}\,,\,\,\,\,\,x^*_{np,2} = -1+\frac{n+1}{\ep_2}\,,
\eq 
with $n\in\N$. We deduce that
\beq\eqlabel{MnpepExpansion}
\Mcal^{np}_{\ep_1,\ep_2}=  \frac{1}{4} \sum_{n=1}^\infty \frac{(-1)^n (-Q)^{\frac{n}{\ep_1}}}{n \sin\left(\frac{\pi n}{\beta}\right)\sin\left(\frac{\pi n}{\ep_1}\right)}+\frac{1}{4} \sum_{n=1}^\infty \frac{(-1)^n (-Q)^{\frac{n}{\ep_2}}}{n \sin\left(\pi n \beta\right)\sin\left(\frac{\pi n}{\ep_2}\right)}\,,
\eq
confirming the universal Chern-Simons based results for the non-perturbative completion of the conifold of \cite{KM15} (under redefinition of the K\"ahler parameter).

Note that as integral
$$
\Mcal_{g_s,-g_s}(Q) = \Mcal_{g_s}(Q)\,,
$$
with $\Mcal_{gs}$ as defined in section \ref{GVsecg0}, eq. \req{MintRep}. Clearly, this relation extends to $\Mcal^p$. However, care has to be taken for the non-perturbative expansion $\Mcal^{np}$, as we can not simply substitute $\beta=1$ in \req{MnpepExpansion}. Rather, a proper limiting procedure has to be performed in order to recover the pole enhancement leading to \req{MnpgsExpansion}.\footnote{We thank R. Mkrtchyan for some related explanations.}

In order to make contact with \cite{LV12}, we note that we can express $\Mcal^{np}$ in terms of $\Mcal^{p}$ as
$$
\boxed{
\Mcal^{np}_{\ep_1,\ep_2}(Q) = \Mcal^{p}_{1/\beta,1/\ep_1}\left(-(-Q)^{1/\ep_1}\right)+\Mcal^{p}_{\beta,1/\ep_2}\left(-(-Q)^{1/\ep_2}\right)
}\,,
$$
recovering under the exponential the triple (perturbative) partition function factor structure of \cite{LV12}, thereby rederiving their result for all geometries with vanishing refined Gopakumar-Vafa invariants at $g_L+g_R>0$.

\paragraph{$g_L+g_R\geq 0$}
Let us move on to the general case. We define 
\beq
\begin{split}
&L^{(g_L,g_R)}_{\ep_1,\ep_2}(Q):=\\
&\prod_{k_L=0}^{2g_L}\prod_{k_R=0}^{2g_R}\prod_{k_1,k_2=0}^\infty
\left(1- q^{\frac{g_L+g_R-k_L-k_R-1-2k_1}{2}}t^{\frac{g_L-g_R-k_L+k_R+1+2k_1}{2}}\, Q\right)^{\frac{(-1)^{k_L+k_R}}{k_L!k_R!(2g_L-k_L)!(2g_R-k_R)!}}\,,
\end{split}
\eq
such that
\beq\eqlabel{ZrefViaLgg}
Z_{ref} = \prod_{\bf d} \prod_{g_L=0}^\infty\prod_{g_R=0}^\infty  L^{(g_L,g_R)}_{\ep_1,\ep_2}(Q^{\bf d})^{\frac{(2g_L)!(2g_R)!}{(-4)^{g_L+g_R-1}}\,n_{\bf d}^{(g_L,g_R)}}\,.
\eq
Under taking the logarithm, making use of \req{LogIntRep}, and performing the summations, we have
\beq\eqlabel{LggDef}
\begin{split}
&\mathcal L^{(g_L,g_R)}_{\ep_1,\ep_2}(Q)=\\
& -\frac{(-4)^{g_L+g_R}}{8\pi\ii\,(2g_L)!(2g_R)!}\int_{\Ccal_+}dx\, \frac{\sin^{2g_L}\left(\frac{\pi\ep_-(1+x)}{2}\right)\sin^{2g_R}\left(\frac{\pi\ep_+(1+x)}{2}\right)\Gamma^2(1+x)\Gamma(-x)}{\sin\left(\pi \ep_1(1+x)\right)\sin\left(\pi \ep_2(1+x)\right)\Gamma(2+x)} (-Q)^{1+x}\,.
\end{split}
\eq
Clearly, as integral
\beq\eqlabel{LtoM}
\begin{split}
\mathcal L^{(0,0)}_{\ep_1,\ep_2}(Q)&=\Mcal_{\ep_1,\ep_2}(Q)\,.
\end{split}
\eq
The perturbative poles of \req{LggDef} can be easily evaluated, leading to
\beq\eqlabel{LggPert}
\mathcal L^{(g_L,g_R),\,p}_{\ep_1,\ep_2}(Q)=\frac{(-4)^{g_L+g_R-1}}{(2g_L)!(2g_R)!}\sum_{n=1}^\infty\frac{\sin^{2g_L}\left(\frac{\pi n\ep_-}{2}\right)\sin^{2g_R}\left(\frac{\pi n\ep_+}{2}\right)}{n \sin\left(\pi n\,\ep_1\right)\sin\left(\pi n\,\ep_2\right)}\,Q^n\,.
\eq
With \req{ZrefViaLgg}, we indeed reproduce the original expansion \req{RefGV} from the perturbative poles.

The non-perturbative pole structure of $\mathcal L^{(g_L,g_R)}$ depends on the values taken by $\ep_i$. Besides possible enhancement to order 2 poles, we may have as well for $g_L+g_R>0$ partial or complete absence of non-perturbative poles. Here, we consider $\beta\not\in \mathbb Q$, such that we have the two rays of non-perturbative poles \req{refNPpoles} for all $g_L,g_R$. Taking residue, yields
\beq\eqlabel{LggNP}
\begin{split}
&\mathcal L^{(g_L,g_R),\,np}_{\ep_1,\ep_2}(Q)\\
&=\frac{(-4)^{g_L+g_R-1}}{(2g_L)!(2g_R)!}\sum_{n\,{\rm odd}}^\infty\left(\frac{\cos^{2(g_L+g_R)}\left(\frac{n\pi }{2\beta}\right)}{n\sin\left(\frac{n\pi}{\beta}\right)\sin\left(\frac{n\pi}{\ep_1}\right)} \, (-Q)^{\frac{n}{\ep_1}}- \frac{\cos^{2(g_L+g_R)}\left(\frac{n\pi \beta}{2}\right)}{n\sin\left(n\pi\beta\right)\sin\left(\frac{n\pi}{\ep_2}\right)} \,(-Q)^{\frac{n}{\ep_2}}\right)\\
&-\frac{(-4)^{g_L+g_R-1}}{(2g_L)!(2g_R)!}\sum_{n\,{\rm even}}^\infty\left(\frac{\sin^{2(g_L+g_R)}\left(\frac{n\pi }{2\beta}\right)}{n\sin\left(\frac{n\pi}{\beta}\right)\sin\left(\frac{n\pi}{\ep_1}\right)} \, (-Q)^{\frac{n}{\ep_1}}-\frac{\sin^{2(g_L+g_R)}\left(\frac{n\pi \beta}{2}\right)}{n\sin\left(n\pi\beta\right)\sin\left(\frac{n\pi}{\ep_2}\right)} \,(-Q)^{\frac{n}{\ep_2}}\right)\,.
\end{split}
\eq

We infer via \req{ZrefViaLgg} that the non-perturbative completion of the refined Gopakumar-Vafa expansion \req{RefGV} for $\beta\not\in\mathbb Q$, necessary for our analytic continuation, reads
$$
\boxed{
\Fcal_{ref}^{np} = \sum_{\bf d}\sum_{g_L,g_R=0}^\infty \frac{(2g_L)!(2g_R)!}{(-4)^{g_L+g_R-1}}\, n_{\bf d}^{(g_L,g_R)}\, \mathcal L^{(g_L,g_R),\,np}_{\ep_1,\ep_2}(Q^{\bf d})
}\,.
$$
A similar remark as before regarding specialization of equivariant parameters applies.

\paragraph{NS limit}
The Nekrasov-Shatashvili limit (for short NS) is defined as \cite{NS09}
$$
\Fcal_{NS}:= \lim_{\ep_2\rightarrow 0}\ep_2 \,\Fcal_{ref}\,.
$$
We can apply the limit to \req{LggDef}, yielding
\beq\eqlabel{LgNS}
\mathcal L^{(g_L,g_R)}_{NS,\,\hbar}(Q)=\frac{(-4)^{g_L+g_R}}{8\pi^2\ii\,(2g_L)!(2g_R)!}\int_{\Ccal_+} \frac{dx}{1+x}\frac{\sin^{2(g_L+g_R)}\left(\frac{\pi\hbar(1+x)}{2}\right)\Gamma^2(1+x)\Gamma(-x)}{\sin\left(\pi \hbar(1+x)\right)\Gamma(2+x)} (-Q)^{1+x}\,.
\eq
Alternatively, we may also directly act with the limit onto the previous results obtained via taking residue (for proper choice of $\beta$). For instance, acting on \req{LggPert} yields
$$
\mathcal L^{(g_L,g_R), \,p}_{NS,\,\hbar}(Q) = \frac{(-4)^{g_L+g_R-1}}{\pi (2g_L)!(2g_R)!}\sum_{n=1}^\infty \frac{\sin^{2(g_L+g_R)}\left(\frac{n\pi\,\hbar}{2}\right)}{n^2 \sin\left(n\pi\, \hbar\right)}\,Q^{n}\,.
$$
(We redefined $\hbar:=\ep_1$.) 

As a side remark, note that using \req{LtoM}, we immediately infer that
\beq\eqlabel{MNSp}
\Mcal^{p}_{NS,\,\ep_1}(Q):=\lim_{\ep_2\rightarrow 0} \ep_2 \,\Mcal^p_{\ep_1,\ep_2}(Q) = \frac{1}{4\pi}\sum_{n=1}^\infty\frac{Q^n}{n^2\sin\left(n\pi\ep_1\right)}\,.
\eq
Acting with the derivative in respect to the surviving equivariant parameter, we easily can verify that (as has been already discussed in detail in \cite{H15,KM15})
$$
\Mcal_q^{np}(Q) = -\frac{\partial}{\partial g_s}\left( g_s\, \Mcal^{p}_{NS,\,1/g_s}\left(-(-Q)^{1/g_s}\right) \right)\,.
$$
Now note that in general the Gopakumar-Vafa invariants $n_{\bf d}$ at tree-level are identical in the refined and unrefined case. Therefore, we can conclude that
\beq\label{F0final}
\boxed{
\Fcal^{(0)} = \Fcal^{(0),\,p}-\frac{\partial}{\partial g_s}\left( g_s\, \Fcal_{NS}^{(0,0),\,p}\left(1/g_s;-(-Q)^{1/g_s}\right) \right)
}\,.
\eq
We recognize that the non-perturbative complete $\Fcal^{(0)}$, derived in section \ref{GVsecg0}, takes the form of the conjecture of \cite{HMMO13}. 

The non-perturbative sector requires some more care. Note that in \req{LgNS} we have for $g_L+g_R=0$ non-perturbative poles at
\beq\eqlabel{NSnpPoles}
x^*_{np}= -1+\frac{n+1}{\hbar}\,,
\eq
with $n\in \N$. Hence, 
$$
\Mcal^{np}_{NS,\,\hbar}(Q) = \frac{\hbar}{4\pi} \sum_{n=1}^\infty \frac{(-1)^n(-Q)^{\frac{n}{\hbar}}}{n^2\sin\left(\frac{n\pi}{\hbar}\right)}\,.
$$
Comparing with \req{MNSp} shows that we have the S-dual like relation
\beq\eqlabel{MnsSDrel}
\boxed{
\Mcal^{np}_{NS,\,\hbar}(Q) =\hbar\, \Mcal^{p}_{NS,\,1/\hbar}\left(-(-Q)^{1/\hbar}\right) 
}\,,
\eq
as first observed in \cite{KM15}.

However, for $g_L+g_R>0$ the poles \req{NSnpPoles} with $n$ odd are cancelled against a vanishing nominator in \req{LgNS}, such that
$$
\mathcal L^{(g_L,g_R),\, np}_{NS,\,\hbar}(Q)= -\hbar\frac{(-4)^{g_L+g_R-1}}{\pi (2g_L)!(2g_R)!}\sum_{n\,{\rm odd}}^\infty\frac{(-Q)^{\frac{n}{\hbar}}}{n^2\sin\left(\frac{n\pi}{\hbar}\right)}\,.
$$
This is consistent with applying the NS limit directly to \req{LggNP}.

Hence, we deduce the necessary non-perturbative completion of the perturbative NS free energy, implied by our formalism, to be given by
$$
\boxed{
\Fcal_{NS}^{np}= \frac{\hbar}{\pi} \sum_{\bf d}\left(n_{\bf d}^{(0,0)}\sum_{n\,{\rm even}}\frac{(-Q)^{\frac{n{\bf d}}{\hbar}}}{n^2\sin\left(\frac{n \pi}{\hbar}\right)}-N_{\bf d}\sum_{n\,{\rm odd}}^\infty \frac{(-Q)^{\frac{n{\bf d}}{\hbar}}}{n^2\sin\left(\frac{n \pi}{\hbar}\right)}\right)
}
\,,
$$
with
$$
N_{\bf d}:=\sum_{g_L,g_R=0}^\infty n_{\bf d}^{(g_L,g_R)}\,.
$$
Note that the definition of $N_{\bf d}$ makes sense, as it seems that one has in general only a finite number of non-vanishing invariants $n_{\bf d}^{(g_L,g_R)}$ at fixed degree ${\bf d}$ (\cf, \cite{CKK12}).

\section{Example: Resolved conifold / Chern-Simons on $\S^3$}
\label{ExampleSec}
In order to illustrate that the Mellin-Barnes integral representation is more widely applicable, let us briefly consider $U(N)$ Chern-Simons theory on $\S^3$. 

Note first that the resolved conifold has $\chi=2$, a single K\"ahler modulus and a single non-vanishing Gopakumar-Vafa invariant at genus zero, equal to one. Hence, \req{Z0def} reduces to 
$$
Z_{Coni}(g_s;Q)= M_{g_s}(1)^{-1}\, M_{g_s}(Q)\,.
$$
Let us consider now $U(N)$ Chern-Simons theory on $\S^3$. The partition function reads (see for instance \cite{M04})
$$
Z_{CS}= \delta^{-N/2} e^{-\frac{\ii\pi N(N-1)}{4}} e^{\frac{\ii \pi N(N^2-1)}{6\delta}} \prod_{m=1}^{N-1}(1-q^m)^{N-m}\,,
$$
with $q:=e^{-\frac{2\pi\ii}{\delta}}$ and $\delta:=\kappa+N$ ($\kappa$ is the Chern-Simons coupling constant). As the logarithm of the partition function takes the canonical form \req{Fqualitative}, we can apply the Mellin-Barnes method introduced in the introduction. Namely, we immediately infer that
\beq
\begin{split}
\log Z_{CS} =&-\frac{N}{2}\log\delta -\frac{\ii\pi(N(N-1))}{4}+\frac{\ii\pi N(N^2-1)}{6\delta} \\
&+ \frac{1}{8\pi \ii}\int_{\Ccal_+} \frac{((q^{N(1+x)}-1)-N(q^{1+x}-1))\,\Gamma^2(1+x)\Gamma(-x)}{\sin^2\left(\frac{\pi(1+x)}{\delta}\right)\Gamma(2+x)}(-1)^{1+x} dx\,.
\end{split}
\eq
Comparing with \req{MintRep} shows that
\beq
\begin{split}
\log Z_{CS} =&-\frac{N}{2}\log\delta -\frac{\ii\pi(N(N-1))}{4}+\frac{\ii\pi N(N^2-1)}{6\delta} \\
 &-\Mcal_{1/\delta}(1)+\Mcal_{1/\delta}(q^N)+\frac{N}{2\pi\ii}\int_{\Ccal_+}\frac{\Gamma^2(1+x)\Gamma(-x)}{(1-q^{-(1+x)})\Gamma(2+x)}(-1)^{1+x} dx\,.
\end{split}
\eq
It remains to evaluate the left-over integral. The perturbative poles at $x^*=n$ lead to a contribution
$$
N\sum_{n=1}^\infty \frac{1}{n(1-q^{-n})}=N\log\phi(q)\,,
$$
where we introduced the logarithm of the Euler function $\phi(q)$, a modular form related to the Dedekind $\eta$-function through a Ramanujan identity as
$$
\phi(q) = q^{-\frac{1}{24}}\eta(\tau)\,,
$$
with $q=e^{2\pi\ii\, \tau}$.

The non-perturbative poles at $x^*= -1+\delta(n+1)$ lead to a contribution 
$$
-N\sum_{n=1}^\infty \frac{1}{n\left(1-q^{n \delta^2}\right)}=-N\log \phi\left(q^{-\delta^2}\right)\,.
$$
Note that via making use of the functional equation of the $\eta$-function, $\eta(-1/\tau)= \sqrt{-\ii \tau}\,\eta(\tau)$, we have the identity
$$
\phi\left(q^{-\frac{1}{\tau^2}}\right)=q^{\frac{1}{24 \tau^2}}\,\eta(-1/\tau)=\sqrt{-\ii \tau}\, q^{\frac{1}{24}\left(1+\frac{1}{\tau^2}\right)} \, \phi(q)\,.
$$
Setting $\tau=-1/\delta$ we conclude that
$$
\boxed{
\log Z_{CS} = \log Z_{Coni}(1/\delta; q^N)-\frac{\ii\pi N^2}{4}+\frac{\ii\pi N(N^2-1)}{6\delta}-\frac{\ii\pi N}{12}\left(\frac{1}{\delta}+\delta\right)
}
\,.
$$
This is an exact statement, as it does not involve a large $N$ limit, nor any other kind of perturbative expansion, confirming the more general discussion of \cite{M13,KM15} in a simple and straight-forward manner. Of course, other gauge groups and the refined case can be discussed similarly.

\acknowledgments

The work of D.K. has been supported in part by the National Research Foundation of
Korea, Grant No. 2012R1A2A2A02046739.

\end{document}